# Islanded Microgrid Restoration Studies with Graph-Based Analysis

Ogbonnaya Bassey, *Member, IEEE* and Karen L. Butler-Purry, *Fellow, IEEE*

*Abstract*—The need to restore and keep the grid running or fast restoration during emergencies such as extreme weather conditions is quite apparent given the reliance of other infrastructure on electricity. One promising approach to electricity restoration is the use of locally available energy resources to restore the system to form isolated microgrids. In this paper, we present a black start restoration method that forms islanded microgrids after a blackout. The master DGs in the formed microgrids are coordinated to work together through droop control. Several constraints, including incentive-based demand response (DR) with direct load control (DLC) and distributed generator (DG) operation constraints, were formulated and linearized to realize a mixed-integer linear programming (MILP) restoration model. To improve compactness and to ensure that the model is neither under-sized nor over-sized, a pre-processing graph analysis approach was introduced which helps to characterize the least number of restoration steps needed to optimally restore the microgrid. Studies were performed on a modified IEEE 123 node test feeder to evaluate the effects of demand response, non-dispatchable DGs, and choice of restoration steps on the quality of the restoration solution.

## Nomenclature

### A. Sets

| | |
|---|---|
| $n(A)$ | The number of elements in set A |
| $B, B^S, B^F, C$ | Set of branches, switchable and damaged branches, set of switchable branches between bus blocks |
| $G, G^{BS}, G^{Dr}, G^F, G^{PQ}$ | Set of all DERs, subset of black start DGs, subset of droop-controlled DGs, subset of damaged DGs, subset of PQ DGs |
| $L, L^S, L^C, L^F$ | Set of loads, subset of switchable loads, subset of controllable loads (with demand response) and subset of damaged loads |
| $T$ | Set of time steps $\{1, 2, \ldots, N_T\}$ and $n(T) = N_T$ |
| $N_p, N, K$ | Set of phase nodes, nodes, bus blocks, $n(N_p) \geq n(N) \geq n(K)$ |

### B. Binary Decision Variables (1 – Energized, 0 – Not Energized)

| | |
|---|---|
| $\hat{x}_{i,t}^N, \hat{x}_{j,t}^K$ | Energization status of node $i$ at time step $t$, energization status of bus block $j$ at time step $t$ |
| $\hat{x}_{g,t}^G$ | Energization status of DG $g$ at time step $t$ |
| $\hat{x}_{ij,t}^{BR}, \hat{x}_{ij,t}^K$ | Energization status of line $(i,j)$ at time step, $t$, where $(i,j) \in B, C$ respectively |
| $\hat{x}_{l,t}^L$ | Energization status of load $l$ at time step, $t$ |

### C. Continuous Decision Variables

| | |
|---|---|
| $a, b,$ or $c$ | Used as subscript or superscript to denote variable or parameter in phase a, b or c, respectively |
| $\hat{P}_{ph,k,t}^{dg}, \hat{Q}_{ph,k,t}^{dg}$ | Active and reactive power output of PQ DG $k$ at step $t$ and phase $ph \in \{a, b, c\}$ |
| $\hat{V}_{n,t}^{re}, \hat{V}_{n,t}^{im}$ | Real and imaginary part of three-phase nodal voltage vector of node $n$ at step $t$ |
| $\hat{I}_{n,m,t}^{re}, \hat{I}_{n,m,t}^{im}$ | Real and imaginary part of three-phase current vector flowing from node $n$ to $m$ at step $t$ |
| $\hat{I}_{n,t}^{re}, \hat{I}_{n,t}^{im}$ | Real and imaginary part of three-phase current vector injected into node $n$ at step $t$ |
| $n_{v,g,t}$ | Voltage droop co-efficient of DG $g$, $g \in G^{Dr}$ at step $t$ |
| $n_{f,g,t}$ | Frequency droop co-efficient of DG $g$, $g \in G^{Dr}$ at step $t$ |
| $\hat{P}_{ph,k,t}^{ref}, \hat{Q}_{ph,k,t}^{ref}$ | Droop reference active and reactive power output of DG $k$ at step $t$, phase $ph$, $k \in G^{Dr}$ |
| $\hat{P}_{ph,l,t}, \hat{Q}_{ph,l,t}$ | Nominal active and reactive power demand of load $l$, phase $ph$, at time step $t$ |

*D. Parameters*

| | |
|---|---|
| $M$ | A large number chosen deliberately to manipulate the constraint equations |
| $\Delta t$ | time interval between restoration steps and is assumed to be a constant value for all intervals |
| $P_g^{G,ramp}, Q_g^{G,r}$ | Maximum absolute value of differential active and reactive power output of DG $g$ for each time step (DG ramp rate) |
| $P_g^{min}, P_g^{max}$ | Minimum and maximum active power output of DG $g$ |
| $Q_g^{min}, Q_g^{max}$ | Minimum and maximum reactive power output of DG $g$ |
| $P_{ph,l,t}, Q_{ph,l,t}$ | Nominal active and reactive power value of load $l$, phase $ph$, at time step $t$ (fixed to the same value for all time steps and is independent of whether the load has been restored or not) |
| $z_{n,k} = r_{n,k} + jx_{n,k}$ | Impedance of line between nodes $n$ and $k$, and $y_{n,k} = \frac{1}{z_{n,k}} = g_{n,k} + jb_{n,k}$ |
| $y_{n,k}^{sh} = g_{n,k}^{sh} + jb_{n,k}^{sh}$ | Shunt admittance between nodes $n$ and $k$ |
| $P_{ph,k,t}^{fc}, Q_{ph,k,t}^{fc}$ | Active and reactive power output from forecast of non-dispatchable PQ DG $k$ at step $t$ and phase $ph$ |

I. INTRODUCTION

A microgrid is an interconnection of distributed energy resources (DERs) and loads coordinated by central and/or decentralized controllers for efficient system operation. A microgrid can operate as a grid-connected controllable system or as an isolated/island system. The ability of the microgrid to operate in island mode during emergency situations can be leveraged for the restoration of distribution systems and for supplying critical loads.

The operation of a microgrid in island mode is particularly challenging due to its relatively smaller system size, uncertainties introduced by distributed energy resources (DERs) penetration, unbalance conditions, and inability to use traditional modeling approaches [1]. Even more, the operation of microgrids during emergency situations such as black start restoration further exacerbate these challenges.

Distribution systems restoration by partitioning the system into microgrids has been widely studied [2-8]. A mixed-integer nonlinear programming (MINLP) approach which considers power loss in its microgrid restoration model was introduced in [9]. More recent distribution system restoration models consider the integration of repair crews/mobile battery-carried vehicles [10] and internal combustion engine cars [11].

While partitioning the distribution system during an emergency can be beneficial in restoring critical loads when parts of the system are damaged, most of the distribution system restoration studies have partly favored this partitioning because the control required to coordinate multiple master distributed generators (DGs) were not considered [2-8]. By coordinating the multiple master DGs to reduce unnecessary partitioning, a bigger system can be restored. Arguably, such a bigger system can improve the resilience of the islanded microgrid simply by not relying on the continuous operation of one master DG but rather multiple cooperating master DGs, by increasing redundancy, and improving phase load balancing. Coordinating multiple master DGs to regulate frequency and voltage together adds more layers of complexity in islanded microgrid operation which, as previously highlighted, is already challenging. Voltage and frequency regulation for microgrid island operation is usually accomplished through droop control [12-15]. Systematic restoration formulation for microgrids operating in droop-controlled island mode is scarce as most existing literature have approached this restoration problem through a rule-based approach [16-19]. This paper aims to bridge the gap by proposing a systematic restoration formulation that considers the coordination of multiple master DGs using droop-control.

Solving sequential restoration problems can quickly become computationally daunting as the number of variables and constraints increase. In [5], a rolling horizon approach was used to break the restoration steps into smaller time horizons to reduce computation time at the cost of possibly realizing a suboptimal solution. In solving the sequential restoration of distribution systems and microgrids, it is usually uncertain as to how many restoration steps or the number of sequences is needed to optimally solve the problem without increasing the computation burden unnecessarily. To minimize the guesswork of choosing solution time steps, a graphical pre-processing approach is introduced in this paper to determine the least number of necessary restoration steps needed to optimally restore the system. Specifically, we introduce what we call the restoration step radius (which determines the conservative step estimate) and the restoration step diameter (which determines the generous step estimate). Increasing the number of restoration steps increases the number of model variables to be defined (since variables have to be defined for every restoration step) which in turn increases the model size and reduces compactness. This graphical evaluation can ensure that the system model is neither over-sized nor under-sized, and keep the restoration problem as compact as possible.

The main contribution of this paper is to propose a systematic formulation for the black start of islanded droop-controlled microgrids considering incentive-based demand response, non-dispatchable DGs, and a graphical pre-processing method to improve model compactness. Apart from our earlier work in [20], no other paper has developed a systematic model for the black start restoration of islanded droop-controlled microgrids to the best of the authors' knowledge. This paper expands our previous work by extending the restoration formulation to include graph analysis, demand response, and non-dispatchable DG considerations.

This paper is organized as follows. Section II consists of a concise introduction to the droop control and reference operation of islanded microgrids. Section III consists of the mathematical formulation of the restoration problem for islanded microgrids. In section IV, a pre-processing graph analysis approach to determine the estimated number of restoration steps is presented. Section V consists of case studies and discussions. Conclusions are presented in section VI.

## II. DROOP REFERENCE OPERATION

Given a set of inverters in a microgrid, the equations describing the conventional droop control of the $i^{th}$ inverter are given as [12],

$$f_i = f^{ref} - n_{f,i}(P_i - P_i^{ref}) \quad (1)$$

$$|\bar{V}_i| = |\bar{V}_i|^{ref} - n_{v,i}(Q_i - Q_i^{ref}) \quad (2)$$

Where $f_i$ is the output frequency of the $i^{th}$ inverter, $f^{ref}$ is the reference frequency in Hz assumed to be fixed and set to the nominal value for all droop inverters, $P_i^{ref}$ is the reference active power in per unit, $P_i$ is the active power output in per unit; $n_{f,i}$ is the frequency droop coefficient in Hz per unit power, $|\bar{V}_i|^{ref}$ is the reference voltage in per unit, $Q_i^{ref}$ is the reference reactive power in per unit, $|\bar{V}_i|$ is the output voltage in per unit, and $n_{v,i}$ is the voltage droop coefficient in per unit power.

If the droop settings are controlled such that $P_i = P_i^{ref}$, then $f_i = f^{ref}$, and such that $Q_i = Q_i^{ref}$, then $|\bar{V}_i| = |\bar{V}_i|^{ref}$. In this work, these two droop operation conditions are termed 'reference power operation', since the active and reactive power reference settings are chosen to match the demanded output power of the inverter. By dispatching the power using this approach, we can eliminate the droop coefficients from the optimization formulation to keep it linear. We assume that the optimal control of the droop settings for the inverters are periodically reset to follow this reference power operating conditions and by so doing nullify the effect of droop coefficients at reference operation. Nevertheless, the droop coefficients still determine the frequency, power, and voltage characteristics of the inverter given any slight deviation from this reference point which will certainly occur as demonstrated with time-domain simulation in [20]. In the optimization formulation that follows, the control variables for the reference power of the droop-controlled inverters are set for every restoration step to match their actual power output to maintain an approximate reference operation.

## III. BLACK START RESTORATION FORMULATION FOR ISLANDED MICROGRIDS

The system is assumed to be configured as an active distribution network disconnected from the main grid with remote controllable switches (RCSs), conventional droop-controlled inverter-based DGs (grid-supporting DGs), dispatchable PQ inverter-based DGs (grid-feeding DGs), non-dispatchable PQ DGs, controllable (demand response) loads, and/or switchable/non-switchable aggregated loads. Before the islanded microgrids are restored, we assume that an unforeseen emergency leads the system to be in a blackout state, and thus, all resources in the system are de-energized to the OFF state. In this completely de-energized state, the system would need to be black-started. The symbols and parameters of the proposed restoration method are shown in the nomenclature section at the beginning of this paper. Decision variables are denoted with a hat.

The flowchart of the proposed restoration model is shown in Fig. 1.

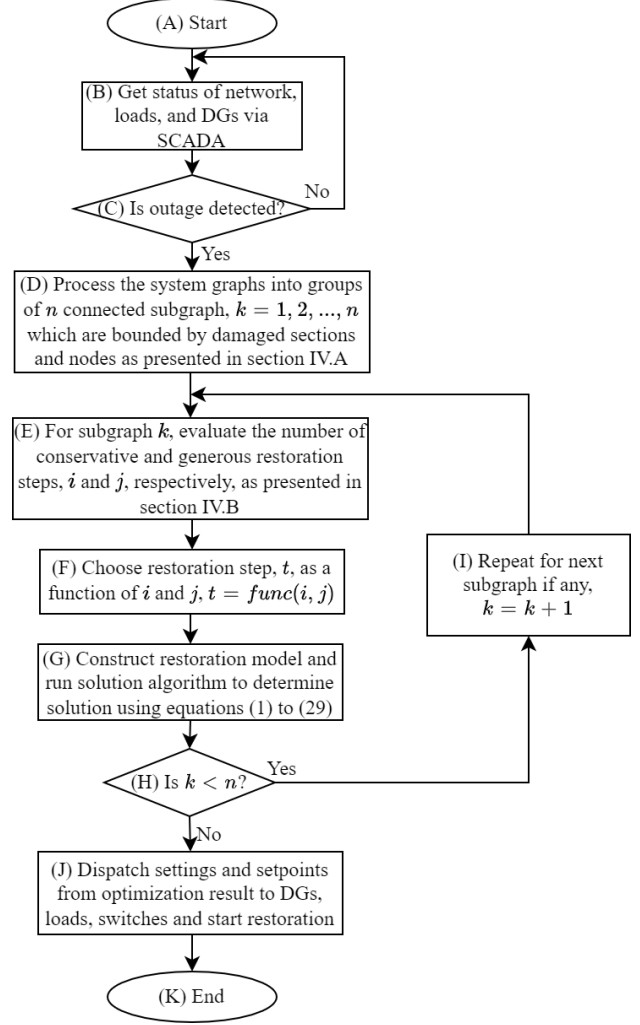

Fig. 1. Flowchart of the proposed restoration method

As shown in the step labeled D on the flowchart of Fig. 1, the distribution system will be divided into a set of connected subgraphs whose boundaries are necessitated by damaged sections and nodes rather than simply by the control requirements of the master DGs as popularly modeled. This way, each subgraph is controlled to be restored as one microgrid system with more redundancy and resilience. Each subgraph is processed and solved as a separate system. The restoration model that follows presents how a typical subgraph is modeled as a MILP problem.

### A. Objective Function

The goal of the objective function is to maximize the total energy restored over the sequence of restoration time steps and is represented mathematically as follows:

$$min - \sum_{t \in T} \sum_{l \in L} \sum_{ph \in \{a,b,c\}} \hat{P}_{ph,l,t} \cdot \Delta t \quad (3)$$

Where $\hat{P}_{ph,l,t} = \hat{x}^L_{l,t} P_{ph,l,t}$. As defined in the nomenclature section, $\hat{x}^L_{l,t}$ is a binary variable which represents the energization status of the load $l$ at time step $t$, and $P_{ph,l,t}$ is the nominal value of load $l$ at time step $t$ for phase $ph$, $\Delta t$ is the time interval between steps and is assumed to be a constant value for all intervals. For demand response loads, $\hat{P}_{ph,l,t}$ is defined differently to vary over a range as follows: $\hat{x}^L_{l,t} P^{min}_{ph,l} \leq \hat{P}_{ph,l,t} \leq \hat{x}^L_{l,t} P^{max}_{ph,l}$, where $P^{min}_l / P^{max}_l$ is the minimum/maximum of the range that the demand response load is allowed to vary. The subsections that follow outline the constraints.

### B. Initial Sequencing Constraints

The initial sequencing constraints ensure that the restoration starts from a feasible node and that the initial states of the system's element are feasible. This is expressed as follows:

$$\sum_{g \in \{G^{Dr}nG^{BS}\}} \hat{x}^G_{g,t} = 1, \sum_{g \in \{G \setminus (G^{Dr}nG^{BS})\}} \hat{x}^G_{g,t} = 0, t = 1 \quad (4)$$

$$\hat{x}^{BR}_{ij,t} = 0, (i,j) \in B^S \setminus B^F, t = 1 \quad (5)$$

$$\hat{x}^{BR}_{ij,t} = 0, \hat{x}^G_{g,t} = 0, \hat{x}^L_{l,t} = 0, (i,j) \in B^F, g \in G^F, l \in L^F, t \in T \quad (6)$$

The first summation in (4) ensures that one black start droop-controlled DG is started as the build-up node and the second summation ensures that all other DG types are disconnected at the first restoration step. To enable the cooperation of multiple master DGs to regulate frequency and voltage, it is assumed that the master DGs are capable of black start and have droop control features. Equation (5) ensures that every switchable branch is de-energized at the first restoration step. Equation (6) ensures that the status of all damaged branches, DGs, and loads are de-energized for all time steps.

### C. Connectivity Constraints

Connectivity constraints ensure feasible interconnection between elements in the system across all restoration steps. Equation (7) ensures that a DG can only connect to a node that has been energized in the previous or same step. Equations (8), (11), and (14) ensure that an energized DG, branch, or load stays energized. Equation (9) ensures that both end nodes of an energized switchable branch are energized. Equation (10) ensures that the status of a non-switchable branch is equal to those of its end nodes. A similar logic used for the branches is established for switchable and non-switchable loads in equations (12), (13) and (14), respectively.

$$\hat{x}^G_{g,t} \leq \hat{x}^N_{g,t}, g \in G, t \in T \quad (7)$$

$$\hat{x}^G_{g,t} - \hat{x}^G_{g,t-1} \geq 0, g \in G, t \in T \quad (8)$$

$$\hat{x}^{BR}_{ij,t} \leq \hat{x}^N_{i,t}, \hat{x}^{BR}_{ij,t} \leq \hat{x}^N_{j,t}, (i,j) \in B^S \setminus B^F, t \in T, i,j \in N \quad (9)$$

$$\hat{x}^{BR}_{ij,t} = \hat{x}^N_{i,t}, \hat{x}^{BR}_{ij,t} = \hat{x}^N_{j,t}, (i,j) \in B \setminus B^S \setminus B^F, t \in T, i,j \in N \quad (10)$$

$$\hat{x}^{BR}_{ij,t} - \hat{x}^{BR}_{ij,t-1} \geq 0, (i,j) \in B^S \setminus B^F, t \in T, i,j \in N \quad (11)$$

$$\hat{x}^L_{n,t} \leq \hat{x}^N_{n,t}, n \in L^S \setminus L^F, t \in T \quad (12)$$

$$\hat{x}^L_{n,t} = \hat{x}^N_{n,t}, n \in L \setminus L^S \setminus L^F, t \in T \quad (13)$$

$$\hat{x}^L_{n,t} - \hat{x}^L_{n,t-1} \geq 0, n \in L^S \setminus L^F, t \in T \quad (14)$$

### D. Power Flow Constraints

We demonstrated how to incorporate novel linear power flow constraints into the restoration model of islanded droop-controlled microgrids without slack bus assumption in our previous work [20, 21]. A more detailed derivation of the linear power flow formulations for islanded droop-controlled microgrid and its optimal power flow extensions have been presented in our previous work [22]. The interested reader can check the aforementioned references for in-depth derivation and analysis; in this section, a summary of how to incorporate them into the restoration problem is described below. The power flow is based on the current injection and is derived from the well-known power flow expression:

$$\bar{Y}\bar{V} = \bar{I} \quad (15)$$

where $\bar{Y}$ is the complex bus admittance matrix, $\bar{V}$ is the complex bus voltage vector, and $\bar{I}$ is the complex current injection vector. The derivation is based on separating (15) into real and imaginary parts, rewriting in expanded form, including the energization status of branches in the admittance matrix terms and finally linearizing the current terms in the right-hand side in terms of nodal voltages in rectangular form. This leads to the following generalized linear power flow constraints in expanded form:

$$\sum_{\substack{k:k \in N \\ k \neq n}} (G_{n,k} \hat{V}^{re}_{k,t} - B_{n,k} \hat{V}^{im}_{k,t}) \hat{x}^{BR}_{n,k,t} + $$
$$\sum_{\substack{k:k \in N \\ k \neq n}} (-G_{n,k} \hat{V}^{re}_{n,t} + \frac{g^{sh}_{n,k}}{2} \hat{V}^{re}_{n,t}) \hat{x}^{BR}_{n,k,t} - \quad (16)$$
$$\sum_{\substack{k:k \in N \\ k \neq n}} (-B_{n,k} \hat{V}^{im}_{n,t} + \frac{b^{sh}_{n,k}}{2} \hat{V}^{im}_{n,t}) \hat{x}^{BR}_{n,k,t} = \hat{I}^{re}_{n,t}(\hat{V}^{re}_{n,t}, \hat{V}^{im}_{n,t})$$

$$\sum_{\substack{k:k \in N \\ k \neq n}} (B_{n,k} \hat{V}^{re}_{k,t} + G_{n,k} \hat{V}^{im}_{k,t}) \hat{x}^{BR}_{n,k,t} + $$
$$\sum_{\substack{k:k \in N \\ k \neq n}} (-B_{n,k} \hat{V}^{re}_{n,t} + \frac{b^{sh}_{n,k}}{2} \hat{V}^{re}_{n,t}) \hat{x}^{BR}_{n,k,t} + \quad (17)$$
$$\sum_{\substack{k:k \in N \\ k \neq n}} (-G_{n,k} \hat{V}^{im}_{n,t} + \frac{g^{sh}_{n,k}}{2} \hat{V}^{im}_{n,t}) \hat{x}^{BR}_{n,k,t} = \hat{I}^{im}_{n,t}(\hat{V}^{re}_{n,t}, \hat{V}^{im}_{n,t})$$

where $G_{n,k} = -g_{n,k}$ ($g_{n,k}$ is a matrix that represents the real part of the branch admittance between nodes $n$ and $k$) and $B_{n,k} = -b_{n,k}$ ($b_{n,k}$ is a matrix that represents the imaginary part of the branch admittance between nodes $n$ and $k$). Note that $\hat{V}^{re}_{n,t} = [\hat{V}^{re,a}_{n,t} \ \hat{V}^{re,b}_{n,t} \ \hat{V}^{re,c}_{n,t}]^T$, $\hat{V}^{im}_{n,t} = [\hat{V}^{im,a}_{n,t} \ \hat{V}^{im,b}_{n,t} \ \hat{V}^{im,c}_{n,t}]^T$, $\hat{I}^{re}_{n,t}$ and $\hat{I}^{im}_{n,t}$ are similarly defined as three-phase current vectors.

To specify the linear power flow constraints for the microgrid to be restored, (16) and (17) have to be defined as constraints for every node and every step of the restoration. The right-hand sides, which are the current injection terms, present some additional linearization tasks for various injection elements and these have been covered in detail in the authors' previous work [20-22].

The shunt admittance of the lines has been ignored when specifying the constraints for each restoration step. Because the distribution lines are relatively shorter compared to bulk power system transmission lines, the shunt admittance can be ignored without incurring significant errors. This is similar to the linear power flow approach used in DistFlow in which the shunt admittance is ignored [23, 24]. The reason for ignoring the shunt admittance is because the relatively small value of the shunt admittance can increase numerical instability especially for relatively larger systems and pose tolerance issues for the optimization solver.

In the above power flow formulation, the nodal voltages are the state variables and would be returned when the optimization is solved. To make the optimization return the current flowing through any distribution line, the following auxiliary constraints can be added for each line of interest (this is essentially Ohm's law in rectangular form):

$$\begin{bmatrix} \hat{I}_{n,m,t}^{re} \\ \hat{I}_{n,m,t}^{im} \end{bmatrix} = \begin{bmatrix} g_{n,m} & -b_{n,m} \\ b_{n,m} & g_{n,m} \end{bmatrix} \begin{bmatrix} \hat{V}_{n,t}^{re} - \hat{V}_{m,t}^{re} \\ \hat{V}_{n,t}^{im} - \hat{V}_{m,t}^{im} \end{bmatrix} \quad (18)$$

The magnitude of the current flow can be constrained as follows based on the line's per-phase ampacity limit, $I_{n,m,max}^{re,ph}$.

$$\left(\hat{I}_{n,m,t}^{re,ph}\right)^2 + \left(\hat{I}_{n,m,t}^{im,ph}\right)^2 \leq \left(I_{n,m,max}^{re,ph}\right)^2 \quad (19)$$

Equation (19) can be linearized by approximating the constraining circle with the sides of a convex polygon [25].

*E. DG Output Constraints*

The active and reactive power limit constraints for each of the droop controlled DGs, $k \in G^{Dr}$, can be expressed as follows:

$$\hat{x}_{g,t}^{G} P_k^{min} \leq \sum_{ph \in \{a,b,c\}} \hat{P}_{ph,k,t}^{ref} \leq \hat{x}_{g,t}^{G} P_k^{max} \quad (20)$$

$$\hat{x}_{g,t}^{G} Q_k^{min} \leq \sum_{ph \in \{a,b,c\}} \hat{Q}_{ph,k,t}^{ref} \leq \hat{x}_{g,t}^{G} Q_k^{max} \quad (21)$$

For dispatchable DGs operating in PQ mode, (20) and (21) are modified to the following equations for the PQ DGs (the PQ DGs are assumed to be single-phase in this formulation).

$$\hat{x}_{g,t}^{G} P_k^{min} \leq \hat{P}_{ph,k,t}^{dg} \leq \hat{x}_{g,t}^{G} P_k^{max}, t \in T, ph \in \{a,b,c\} \quad (22)$$

$$\hat{x}_{g,t}^{G} Q_k^{min} \leq \hat{Q}_{ph,k,t}^{dg} \leq \hat{x}_{g,t}^{G} Q_k^{max}, t \in T, ph \in \{a,b,c\} \quad (23)$$

For non-dispatchable renewable DGs operating in PQ mode, a simple approach of fixing their power output to its forecasted values, $P_{ph,k,t}^{fc}$ and $Q_{ph,k,t}^{fc}$, when they are energized is adopted. From the time-domain simulation performed for inverter-based droop-controlled restoration in [20], we see that the transient response of this sort of inverter-based DGs are quite fast and thus, the time interval between the sequence of restoration steps can be set in the order of a few seconds. Because of how fast this sort of restoration can be completed, we can reasonably assume the output is equal to their forecasted value without involving in-depth stochastic analysis. Thus, their constraints are expressed as follows:

$$\hat{P}_{ph,k,t}^{dg} = \hat{x}_{g,t}^{G} P_{ph,k,t}^{fc} \quad (24)$$

$$\hat{Q}_{ph,k,t}^{dg} = \hat{x}_{g,t}^{G} Q_{ph,k,t}^{fc} \quad (25)$$

*F. Demand Response Loads Constraints*

Demand response (DR) loads are loads that can participate in demand response and can be varied over the time horizon of interest. The demand response model utilized was the incentive-based DR with direct load control (DLC) program [26, 27]. In this sort of model, the customers get some form of rewards/incentives for agreeing to let the microgrid central controller (MGCC) and/or system operator directly and remotely control their loads during an emergency. The DLC program enables the MGCC to directly control the DR loads for each time step. This direct control of loads can be considered the most suitable form of demand response for black start restoration since the loads can be varied quickly by the MGCC to meet generation balance and phase balancing needs without customer interference. The following constraints are included for each controllable demand response load:

$$\hat{x}_{l,t}^{L} P_l^{min} \leq \hat{P}_{ph,l,t} \leq \hat{x}_{l,t}^{L} P_l^{max}, l \in L^C \quad (26)$$

$$\hat{x}_{l,t}^{L} Q_l^{min} \leq \hat{Q}_{ph,l,t} \leq \hat{x}_{l,t}^{L} Q_l^{max}, l \in L^C \quad (27)$$

$$\frac{\hat{P}_{ph,l,t}}{P_l^{max}} = \frac{\hat{Q}_{ph,l,t}}{Q_l^{max}}, l \in L^C \quad (28)$$

$$\hat{P}_{ph,l,t+1} \geq \hat{P}_{ph,l,t}, \hat{Q}_{ph,l,t+1} \geq \hat{Q}_{ph,l,t}, l \in L^C \quad (29)$$

Equations (26) and (27) ensure that the loads are controlled within their allowable limits and assume a continuous control range. Equation (28) is a constraint that maintains the active and reactive load settings ratio which is equivalent to the assumption that the load's power factor remains the same as it is varied. Equation (29) ensures that the demand response works such that the load curtailment doesn't increase with time step but rather is forced to decrease or leave the curtailment as it was in the previous step; this is to ensure that specific loads within the DR aggregate load stay on after the DLC program commands it to energize. In other words, (29) ensures that any specific load within each aggregate load participating in demand response will stay on after it has been energized instead of flipping on and off over the restoration time steps.

*G. Other Constraints*

Other constraints incorporated in the restoration model include: synchronization enhancing constraints, phase voltage unbalance rate, voltage limit, DG power unbalance, nominal system load unbalance index (NSLUI), topology and sequencing, and ramp rate constraints. These constraints and how they are incorporated into the model have been discussed in our previous work [20, 21]. Synchronization enhancing constraints, for instance, are conditions that can minimize angle, frequency, and voltage transients during the synchronization of droop-controlled master DG to an already operating islanded microgrid. These synchronization constraints ensure that at any restoration step in which a master DG is synchronized to the microgrid, no additional loads are restored and the dispatch settings of all other DGs operating in the system remain as they were in the previous step. Essentially, it means "freezing" the microgrid until the synchronization step is completed.

## IV. IMPROVING COMPUTATION BY GRAPHICAL EVALUATION

In this section, we present some graph-based pre-optimization processing that can ensure that the restoration model is neither over-sized nor under-sized.

*A. Network Graph Evaluation*

A graph is connected if every vertex is joined to every other vertex by a path. A disconnected graph is a graph that is not connected, that is, not every pair of vertices has a path joining them [28].

To evaluate the system topology for restoration, an undirected graph is generated which includes nodes (or vertices) and edges (or branches) with all damaged branches removed. If the resulting graph is connected, then the system data is used as inputs to the restoration algorithm. However, if the graph is disconnected, then it can be grouped into two or more connected subgraphs with each component subgraph solved separately by the restoration algorithm. The graphical analysis described next is performed on each of the subgraphs as a separate independent system to be restored.

*B. Estimating the Number of Restoration Steps*

We first introduce the concept of a bus block, described in [4]. A bus block is a group of nodes connected by non-switchable branches. Grouping the distribution system nodes into a set of bus blocks, $K$, decreases the size of the distribution graph in which edges are represented by a set of switchable branches between bus blocks, $C$. Fig. 2 shows an example of forming a graph from a distribution system using bus blocks.

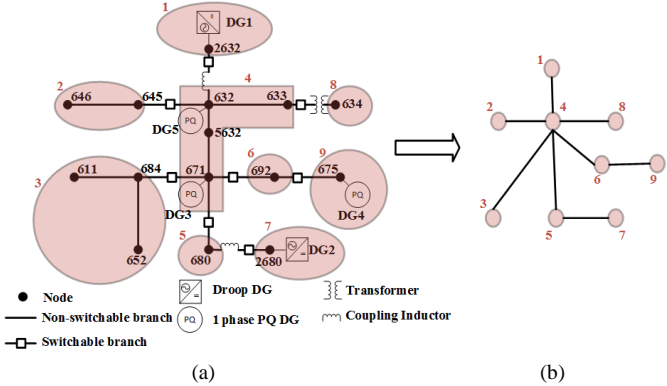

Fig. 2. Illustration of Bus Block Reduction, (a) Modified IEEE 13 Node Test Feeder, (b) Bus Block Equivalent

To estimate the number of time steps required to restore each connected subgraph or entire connected system (if connected), the system is reduced to a set of bus blocks and then we apply the concept of distance and eccentricity from graph theory which is introduced next. Each bus block is regarded as a vertex and each switchable branch between bus blocks is regarded as an edge.

The concept of distance and eccentricity of a vertex/node in a graph presented in [28, 29] is used. Distance and eccentricity are defined as follows.

Distance: Let $u, v$ be vertices in a graph G ($u, v \in V(G)$). The distance from $u$ to $v$ is the length of the shortest path from $u$ to $v$, and is denoted $d(u,v)$ [29].

Eccentricity: The eccentricity, $e(v)$, of a vertex $v$ in a graph $G$ is given by the maximum of all the distances measured from $v$ to every other vertex [29]. That is,

$$e(v) = \max \{d(u,v)|u \in V(G)\} \quad (30)$$

While the distance gives the minimum number of restoration steps required to get from vertex $v$ to an arbitrary vertex $u$, the eccentricity gives the maximum of all the minimum number of restoration steps required to get from vertex $v$ to every other vertex in the graph. Therefore, if the startup node for the restoration is $v$, then the eccentricity gives the minimum number of restoration steps required to get to every vertex in the system.

*1) Restoration Step Diameter and Radius*

The term *restoration step diameter* (RSD), denoted as $RSD(G)$, is defined as the maximum of the eccentricities of vertices representing nodes where black start DGs are connected. Let $V_{BS}(G)$ represent the set of vertices with black start DGs. Then, RSD can be written as,

$$RSD(G) = \max \{e(v)|v \in V_{BS}(G)\} \quad (31)$$

The term *restoration step radius* (RSR), denoted as $RSR(G)$, is defined as the minimum of the eccentricities of vertices representing nodes where black start DGs are connected. RSR can be written similarly as:

$$RSR(G) = \min \{e(v)|v \in V_{BS}(G)\} \quad (32)$$

The RSD gives a generous estimate of the number of time steps required to get from any of the black start vertices to all other vertices in the system. The RSR gives a conservative estimate of the number of time steps required to get from the black start nodes with minimum eccentricity to all other vertices in the system.

*2) Generous and Conservative Restoration Steps Estimates*

Assuming that the ramp rates of the DGs are sufficiently high, then a generous estimate for the required time steps for the solution method is given as,

$$n_g(T) = RSD(G) + n(G^{BS}) \quad (33)$$

The second term of (33), $n(G^{BS})$, is the number of black start DGs available for restoration and is added to account for the zero-dispatch synchronization steps when restoration of the branches and loads are paused temporarily for each of the droop-controlled black start DGs to smoothly synchronize.

Similarly, a conservative estimate for the required time steps is calculated by replacing RSD in (33) with RSR,

$$n_c(T) = RSR(G) + n(G^{BS}) \quad (34)$$

The significance of the RSD and RSR in solving for a restoration solution will be discussed in the case studies section that follows.

V. CASE STUDIES AND DISCUSSION

In this section, we present case studies, performance evaluation, and discussions based on an islanded microgrid adapted from the IEEE 123 node test feeder [30]. A base case study that highlights the key features of the developed method is first presented, followed by derivative performance studies considering non-dispatchable renewable DG sizing, the flexibility of the demand response loads, and choice of restoration steps.

*A. Description of the Base Test System*

A one-line diagram of the base test system is shown in Fig. 3. The system is assumed to have experienced a blackout due to an unforeseen emergency that disconnects the distribution system from the bulk grid. Without loss of generality, we have chosen the case system of Fig. 3 to be a connected graph. If they are not connected, they can be processed into subgraphs and the same restoration formulation can be used to solve each connected subgraph independently as highlighted previously in section IV and the flow chart shown in Fig. 1.

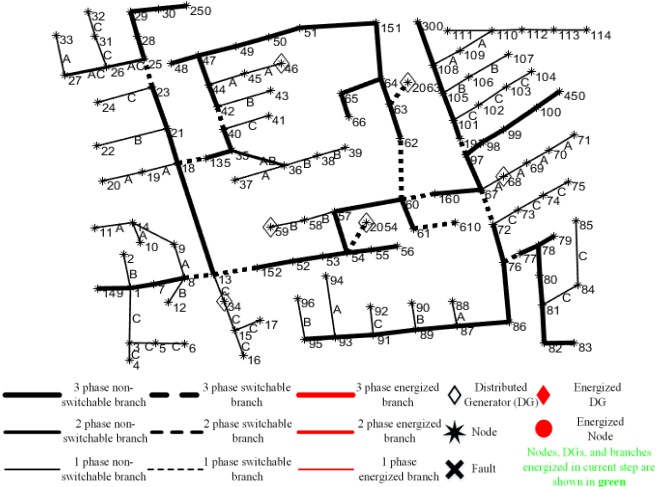

Fig. 3. One-line diagram of the base test system before initiating restoration based on modified IEEE 123 node test feeder

Initially all of the elements are set in a de-energized state which is why the elements are shown in black and white for now in Fig. 3. Two three-phase droop-controlled DGs are present at nodes 2054 and 2063 (these two nodes are extra nodes added to the IEEE 123 node test system to represent the additional nodes due to the inductor coupling of the droop DGs which help to decouple the active and reactive power control). Three single-phase PQ dispatchable DGs are situated at nodes 34, 46, and 59, and a single-phase non-dispatchable renewable DG is situated at node 68. The details of the DGs are shown in Table I. The DGs have a total active power capacity of 2680 KW and a reactive power capacity of 1480 KVAR for all three phases combined. Line ampacity constraints have been ignored.

There are 81 ZIP (constant impedance, current, and power) spot loads. Without loss of generality, the ZIP load co-efficient for every load has been set to 0.4, 0.3, and 0.3 for constant impedance, current, and power components, respectively. 10 out of these 81 ZIP loads are assumed to have demand response capability with full load range controllability, that is, the load can be controlled to operate between its nominal and zero ratings. The sum of the nominal value of loads have a total active power of 3470 KW (1201, 1074, and 1195 KW for phases A, B, and C, respectively), and total reactive power of 1935 kVAR (656.1, 626.5, and 652.4 KVAR for phases A, B, and C, respectively). The value of other components can be found in the IEEE 123 node test feeder [30].

### B. Restoration Solution

The restoration formulation was implemented as a MATLAB program and solved using the Gurobi 9.1.1 optimization solver. The program was solved in a Windows computer with Intel Core i5-7200U CPU @ 2.5GHZ 2.71GHz CPU, 8 GB of RAM, and a 64-bit operating system. The microgrid black start restoration was solved within a solver time of 168.5s with an optimality gap setting of 1% and choice of restoration steps of 7 (as we will discuss in section V.E, this is equal to the generous restoration steps gotten from the pre-processing graph analysis).

The one-line diagram of the restoration sequence showing only the restoration of nodes, branches, and DGs is shown in Fig. 5. Notice a fully interconnected microgrid is restored.

The loads were restored sequentially and coordinated with the single-phase DGs to balance the load unbalance in the system according to the system unbalance constraints. The load restoration results are summarized in Fig. 4 with the total loads restored per step shown with the bars. Notice that in step 4 (Fig. 5) when the second droop-controlled DG is restored, the total loads restored remained the same as the previous step in accordance with the synchronization enhancing constraints.

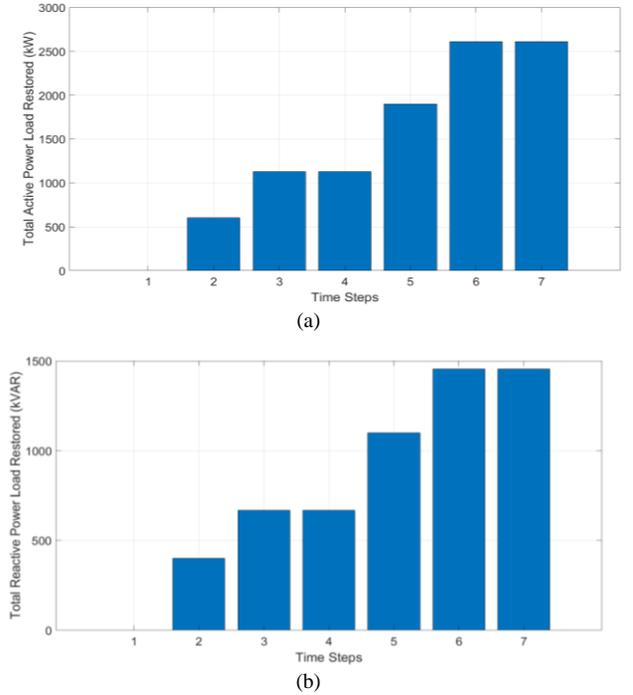

Fig. 4. Total load value restored per restoration steps, (a) active power loads (b) reactive power loads

TABLE I
DG PARAMETERS USED FOR THE CASE STUDY

| LABEL | NODE | TYPE | PER PHASE BASEMVA | PER PHASE BASEKV | PU COUPLING INDUCTOR | PMAX (KW) | PMIN (KW) | QMAX (KVAR) | QMIN (KVAR) | PHASE | WORKING? | BLACKSTART? | RAMP RATE % |
|---|---|---|---|---|---|---|---|---|---|---|---|---|---|
| DG1 | 2054 | DROOP | 1 | 2.4018 | 0.3 | 1200 | 0 | 700 | -160 | ABC | YES | YES | 60 |
| DG2 | 2063 | DROOP | 1 | 2.4018 | 0.3 | 1000 | 0 | 500 | -120 | ABC | YES | YES | 60 |
| DG3 | 34 | PQ | NA | NA | NA | 150 | 0 | 100 | -20 | C | YES | NO | 60 |
| DG4 | 46 | PQ | NA | NA | NA | 130 | 0 | 70 | -15 | A | YES | NO | 60 |
| DG5 | 59 | PQ | NA | NA | NA | 120 | 0 | 70 | -10 | B | YES | NO | 60 |
| DG6 | 68 | PQ (RENEWABLE) | NA | NA | NA | 80 | 80 | 40 | 40 | A | YES | NO | 100 |

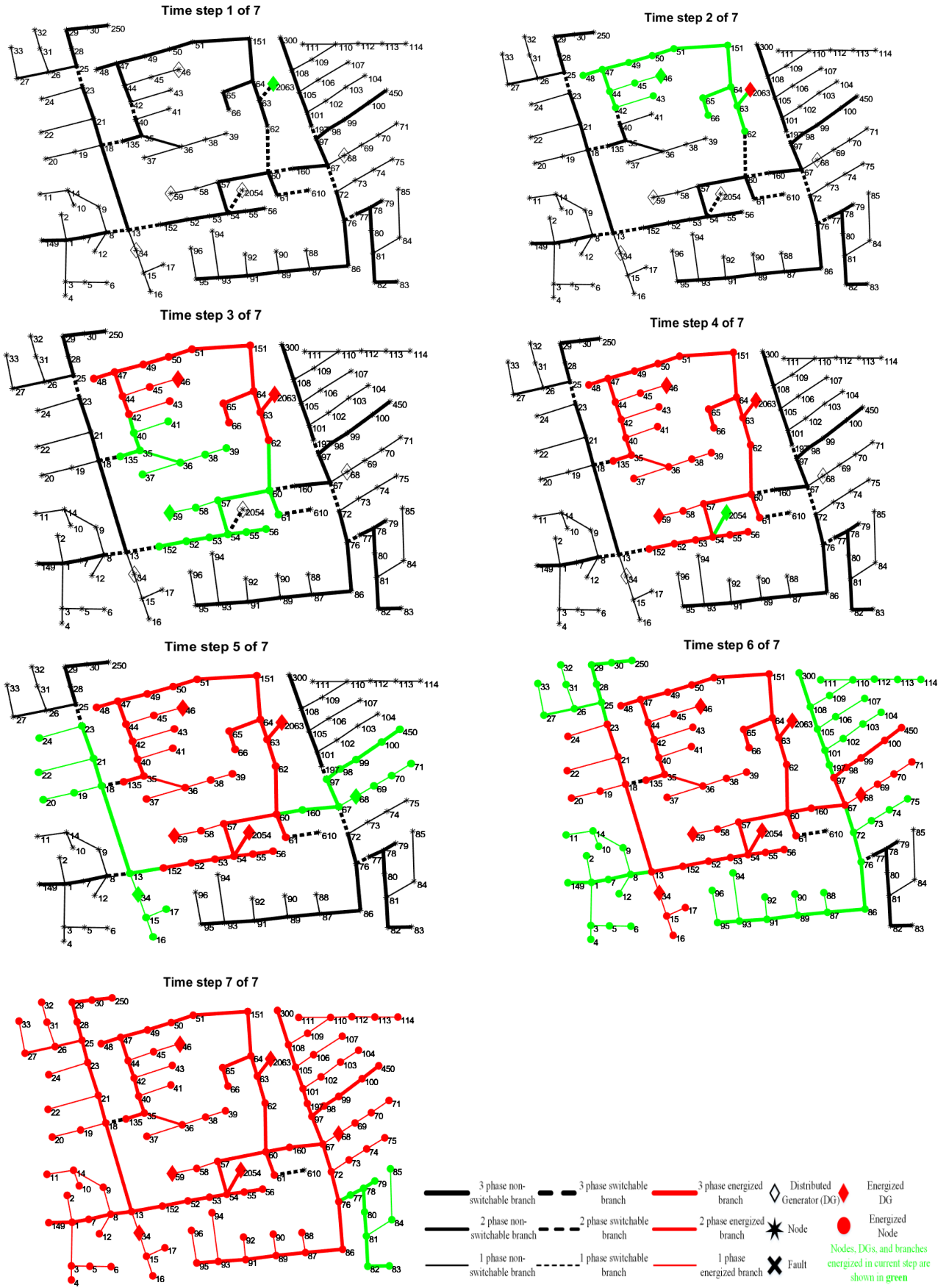

Fig. 5. Restoration sequence of the restored microgrid representing restoration steps 1, 2, to 7 respectively

## C. Non-dispatchable DG Performance Studies

The incorporation of non-dispatchable renewable energy DG in restoration is a major concern because of the intermittent nature of its supply. The performance of the restoration method with non-dispatchable DG is evaluated by varying the output capacity of the non-dispatchable DG of the base case with a factor ranging from 0.25 to 16 and observing how this affects the objective value. The change in objective function magnitude with non-dispatchable DG capacity is shown in Fig. 6.

Notice that as the non-dispatchable DG capacity factor is increased (that is its output value is multiplied by a factor), the objective magnitude (which is the energy restored in KW-steps) increases until a certain capacity threshold where the connection of the non-dispatchable DG is either infeasible or non-beneficial due to phase balancing issues and its non-controllability. This capacity threshold can be roughly estimated from Fig. 6 as a capacity factor of above 8 while the optimal capacity factor is estimated to be around a capacity factor of 4. The optimal and threshold capacity factor can be estimated with higher resolution by running more studies around these regions.

This study suggests that the optimal and threshold capacity factors for a microgrid should be properly studied to determine the size of the non-dispatchable DG that can be installed. Also, incorporating output curtailment for relatively large non-dispatchable DG sources may help in making its connection feasible and beneficial to the islanded microgrid during restoration.

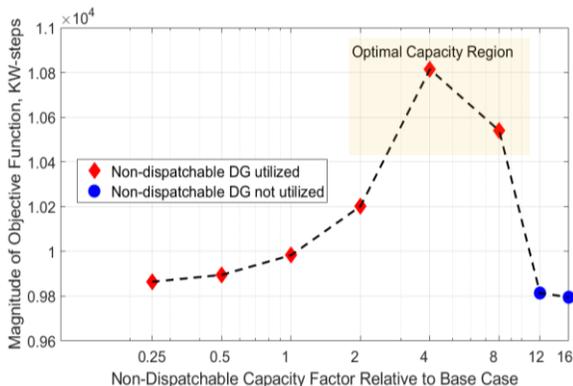

Fig. 6. Estimating Non-dispatchable DG's Optimal and Threshold Capacity under the Same Operating Conditions

## D. Demand Response Loads Performance Studies

To investigate the effect of incorporating demand response, the base case of section V.A is re-run with varying lower limits of load controllability for the 10 loads with demand response. This is summarized in Fig. 7. Generally, we would expect that as the lower bounds of the demand response loads are increased, the energy restored would decrease due to reduced controllability. When the lower bounds factor is set to 1 (that is setting the lower bound equal to the upper bound), direct load controllability is lost and is equivalent to disabling the demand response function. Under this condition, we see a decrease in the restored energy corresponding to the x-axis value of 1 as shown in Fig. 7. It is expected that increasing the number of loads with direct load control will improve the restoration result. In terms of per-phase energy balance in the system, the demand response loads help to improve load balancing across the phases.

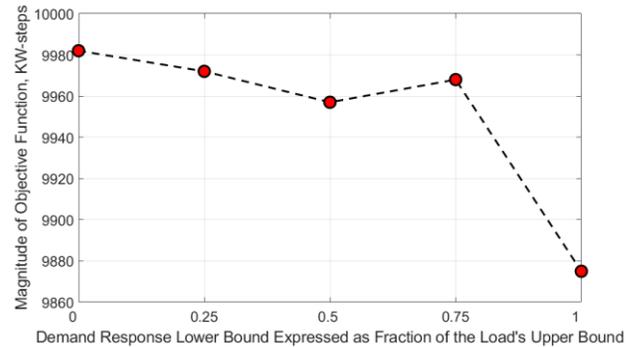

Fig. 7. Effects of Demand Response on the Objective Magnitude (Restored Energy)

## E. Effects of Restoration Steps

The choice of the number of restoration steps is an important parameter in determining if the restoration solution optimally utilizes the available local resources. A choice of a lower number of restoration steps can lead to a sub-optimal solution while a choice of a larger number can lead to a better solution but at the cost of increasing the number of variables and constraints (in other words, decreasing the compactness of the model). To study the effects of the choice of restoration steps on the base case, the graph analysis approach presented in section IV is applied to the base case of the microgrid shown in Fig. 3. The resulting conservative and generous time steps were estimated as 6 and 7, respectively. The base case is then re-solved with varying restoration steps ranging from 4, 5, 6, to 9 time steps. The total active power of load restored at the last step is presented in Fig. 8.

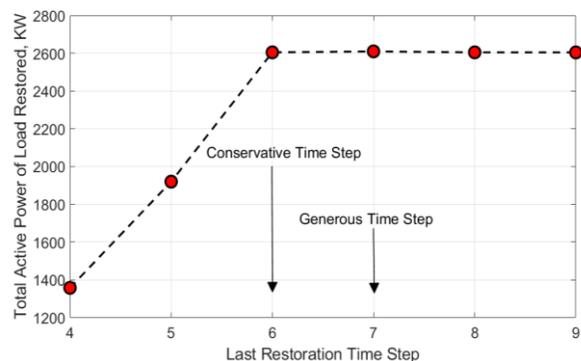

Fig. 8. Total Actve Power of Load Restored using Different Restoration Steps Parameter

Notice that the conservative and generous time steps were sufficient to restore the most loads. Additional studies showed that these time step estimates were not always sufficient to restore the most load and its sufficiency is mostly directly dependent on the ramp rate of the DGs and a complex interplay of other operational constraints. When it is not certain whether these time step estimates are sufficient, the time step parameter can be increased at the cost of reducing model compactness. Nevertheless, computing the conservative and generous time steps can be used to inform the least number of restoration step parameter needed to restore the most load.

The effect of increasing the restoration steps parameter on compactness can be seen in Fig. 9. Notice that the solution time step increases quadratically as the number of variables increases.

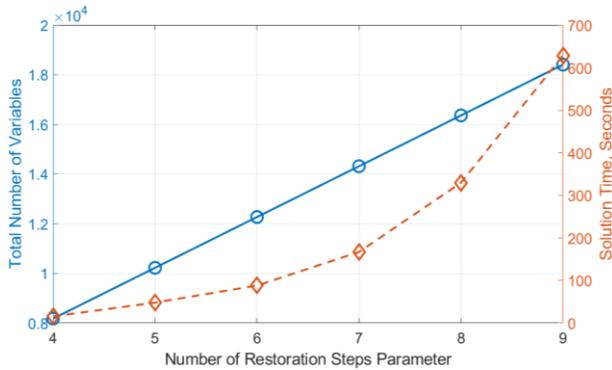

Fig. 9. Effects of Increasing Restoration Steps Parameter on Number of Model Variables and Solution Time

## VI. Conclusion

In this paper, we have presented a black start restoration formulation for islanded microgrids operating in droop mode. The objective of the restoration formulation is to maximize the energy restored over a given restoration step. Several constraints were formulated and linearized to realize a MILP problem. Demand response loads with direct load control and non-dispatchable DG operation were incorporated into the restoration formulation. To improve model compactness, a graph analysis approach was introduced to characterize the restoration step radius and diameter which are then used to compute the conservative and generous time step estimates. These time step estimates help to inform the least number of time steps needed to optimally restore the microgrid. A potential area for future work would be integrating other variants of droop control and considering the stochastic nature of non-dispatchable DER into the restoration formulation.